\documentclass[final,3p,times,twocolumn]{elsarticle}

\usepackage{amssymb}
\usepackage{bm}
\usepackage{amsthm}
\usepackage{lineno}

\journal{Nuclear Instruments \& Methods in Physics Research A}

\begin{document}

\begin{frontmatter}



\title{Development of Transient $\mu$SR Method for High-Flux Pulsed Muons}


\author[KEK,MUSE]{Shoichiro Nishimura}
\ead{shoichiro.nishimura@kek.jp}
\author[KEK,Kinken]{Hirotaka Okabe}
\author[Ibadai]{Masatoshi Hiraishi}
\author[Muroran]{Masanori Miyazaki}
\author[KEK,MUSE]{Jumpei G. Nakamura}
\author[KEK,MUSE,Sokendai]{Akihiro Koda}
\author[KEK,MUSE]{Ryosuke Kadono}
\ead{ryosuke.kadono@kek.jp}
\address[KEK]{Muon Science Laboratory, Institute of Materials Structure Science, High Energy Accelerator Research Organization (KEK),\\ Tsukuba, Ibaraki 305-0801, Japan}
\address[MUSE]{Materials and Life Science Division, J-PARC Center, Tokai, Naka, Ibaraki 319-1106, Japan}
\address[Kinken]{Institute for Materials Research, Tohoku University, Sendai 980-8577, Japan}
\address[Ibadai]{Graduate School of Science and Engineering, Ibaraki University, Mito, Ibaraki, 310-8512, Japan}
\address[Muroran]{Muroran Institute of Technology, Muroran, Hokkaido 050-8585, Japan}
\address[Sokendai]{Department of Materials Structure Science, The Graduate University for Advanced Studies (Sokendai), Tsukuba, Ibaraki 305-0801, Japan}

\begin{abstract}
In order to expand the applicability of muon spin rotation, relaxation, and resonance ($\mu$SR) experiments with pulsed muons and to make effective use of the high-flux beam, we have developed a new experimental method ``transient $\mu$SR''.
In this method, $\mu$SR data for each muon pulse are tagged with external parameters such as temperature and magnetic field in real time, allowing the sample environment to be changed without interrupting data collection. As a result, continuous $\mu$SR measurements under sample conditions that vary on a time scale longer than the beam pulse interval (40 ms) are realized, and efficient beam utilization is achieved by eliminating the lag time associated with moving between discrete measurement points. 
The transient $\mu$SR method was applied to the study of the magnetic properties of cupric oxide (CuO) and the observation of level-crossing resonance relaxation in copper, and the method was successfully established by confirming that the results reproduce those in the previous reports.
\end{abstract}




\end{frontmatter}


\section{Introduction}
\noindent

In recent years, the muon beam intensity provided by the Muon Science Experimental Facility (MUSE) of the Materials and Life Science Experimental Facility (MLF) at the Japan Proton Accelerator Research Complex (J-PARC) has become strong enough to be called high flux ($>$10$^{5}$~$\mu^+$ s$^{-1}$cm$^{-2}$), and the data collection time for a single time spectrum in $\mu$SR experiments has become increasingly short. On the other hand, in conventional $\mu$SR experiments, measurements are performed by setting up discrete measurement points with different conditions such as temperature and magnetic field, and the time required to move to each measurement point with different conditions often greatly exceeds the data collection time, thus slowing down the overall experiment time. In particular, rapidly changing temperature and controlling its disturbances are often difficult and limit the efficiency of experiments; unexpected fluctuation during data collection can interrupt and delay measurements (or reduce data reliability). We have developed a next-generation data acquisition and analysis method to fundamentally solve these difficulties and take advantage of the high-flux beam.

This method integrates $\mu$SR data with information on sample conditions such as temperature and magnetic field in real time and sorts them into a multi-dimensional histogram of positron events for each muon pulse. This enables continuous $\mu$SR measurements under varying sample conditions on a time scale longer than the beam pulse interval (40~ms) and efficient beam utilization is achieved by eliminating the lag time associated with moving between discrete measurement points in conventional $\mu$SR experiments.  It also opens up the possibility of studying transient phenomena that were inaccessible with conventional $\mu$SR. However, for such a method to be practical, it is essential that the beam intensity is high enough to ensure sufficiently good statistics for positron events in the time region of interest. In this respect, the high-flux pulsed muon beam at the MUSE is most suitable for transient $\mu$SR.

Historically, numerous examples are known of attempts to utilize the periodic time structure of an incident beam to efficiently collect data under a time-varying sample environment while minimizing systematic errors. To our knowledge, the oldest example in $\mu$SR is the ``stroboscopic technique'', which performs high-precision spin rotation frequency measurements by synchronizing the data acquisition (DAQ) with the micro-bunch structure of the muon beam reflecting the radio frequency (RF) for the proton acceleration in the cyclotron \cite{Klempt:82,Schenck:85}. Subsequently, the method called ``red-green mode'' has been developed in synchrotron-based muon facilities, where data are collected while irradiating RF/flash lamp/laser pulses or applying a pulsed magnetic field synchronized with the pulsed beam \cite{Ishida:84,Nishiyama:86,Kadono:94,Kadono:03,Ghandi:07,Shimomura:12,Yokoyama:17,Yokoyama:21,Murphy:22,Motokawa:91,Shiroka:99}.  Recently, the red-green mode has been extended in combination with ``event mode'' DAQ, where up to 14~bits (=16384) of different conditions can be pre-programmed for each pulse, enabling the collection and analysis of $\mu$SR and neutron data \cite{Cottrell:12,Giblin:14,PETERSON201524,ARNOLD2014156}. However, the limitation of these measurements is that they can only be applied when the sample environment can be controlled synchronously with the DAQ. The newly developed transient $\mu$SR is novel in that it removes this limitation and can respond to changes in the sample environment that are asynchronous to muon pulse and/or DAQ.

In this paper, we detail the principle of the transient $\mu$SR method and report the results of its application to the study of the magnetic properties of CuO and the observation of muon level-crossing resonance ($\mu$LCR) relaxation in Cu to evaluate its effectiveness.

\begin{figure}[h]
 \centering
  \includegraphics[width=0.42\textwidth,clip]{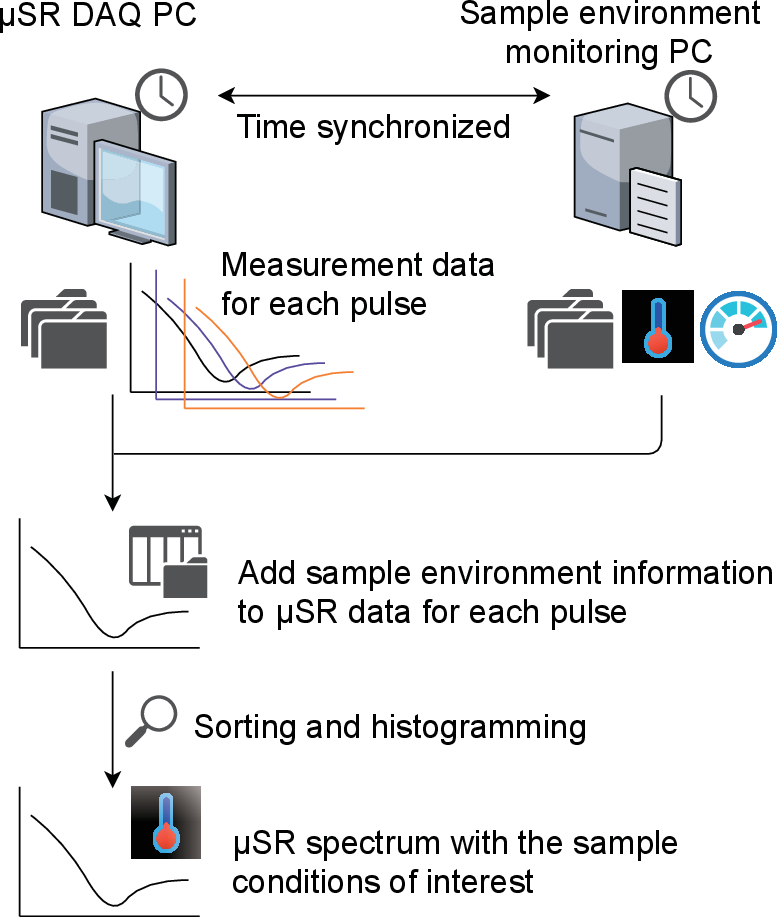}
 \caption{Data acquisition (DAQ) and event analysis flow-chart of the transient $\mu$SR. The PC's for DAQ and sample condition monitoring (SCM) are synchronized by an NTP server that synchronizes the internal clocks. The SCM information is captured in real time by the DAQ PC and recorded/stored with positron event information for each beam pulse. The event-analysis software is invoked for the sorting/histogramming of integrated data. 
}
 \label{Fig1}
\end{figure}

\section{Transient $\mu$SR Measurements}
\noindent
\subsection{Data structure}
\noindent
The purpose of the $\mu$SR experiment is to determine the time evolution of the spin polarization [$G(t)$] exhibited by muons stopped in a sample. This is achieved by monitoring the time-dependent forward-backward asymmetry [$A(t)$] of the positrons emitted from muons with detectors (hodoscopes) and extracting it using the following relation \cite{Yaouanc:00,Blundell:21}, 
\begin{equation}
A(t)= A_0G(t)=\frac{\alpha N_+(t)- N_-(t)}{\alpha N_+(t)+ N_-(t)}\:,\label{asy}
\end{equation}
where $A_0$ is the average asymmetry, 
$$N_\pm(t)= N_{0\pm}e^{-t/\tau_\mu}[1\pm A_\pm G(t)]$$ 
is the time-dependent positron events summed up for the forward ($+$) or backward ($-$) detectors, $\tau_\mu$ is the muon decay lifetime ($=2.197\times10^{-6}$~s), $A_\pm$ is the decay positron asymmetry for the detector in question ($A_\pm\simeq A_0$), and $\alpha$ is the instrumental asymmetry [$\alpha=N_-(0)/N_+(0)$]. 

In conventional $\mu$SR experiments, measurements are made under constant sample conditions such as temperature ($T$), magnetic field ($B$), etc., and information on the local magnetic field in the sample is obtained by analyzing the time spectrum obtained by the Eq.~(\ref{asy}). The new method monitors these sample conditions in real time and acquires the data as multidimensional histograms, namely,
\begin{equation}
\tilde{A}(t;T,B,...)=\frac{\alpha \tilde{N}_+(t)- \tilde{N}_-(t)}{\alpha\tilde{N}_+(t)+ \tilde{N}_-(t)}\:,\label{asyM}
\end{equation}
where
$$\tilde{N}_\pm(t)= N_{0\pm}e^{-t/\tau_\mu}[1\pm A_\pm \tilde{G}(t;T,B,...)].$$
Since $\tilde{G}(t;T,B,...)$ also contains information related to the physical properties of the sample, $\tilde{A}(t)$ is necessarily a measurement quantity with large dimensions.

\subsection{Data Acquisition and Analysis}
As illustrated in Fig.~\ref{Fig1}, integration of $\mu$SR data and information on sample conditions ($T$, $B$, etc.) was achieved by synchronizing the internal clock of the DAQ-PC and another PC for monitoring these conditions using an NTP (Network Time Protocol) server. The integrated data were recorded pulse by pulse, and off-line analysis was performed to extract the data in units of beam pulses by imposing necessary conditions for data selection, and organize them as a multidimensional histogram $\tilde{A}(t)$ that includes information on the sample environment via the Trees in the ROOT data structure. In the following, this first stage of analysis will be referred to as ``event analysis''.

\begin{figure}[ht]
 \centering
\includegraphics[width=0.45\textwidth,clip]{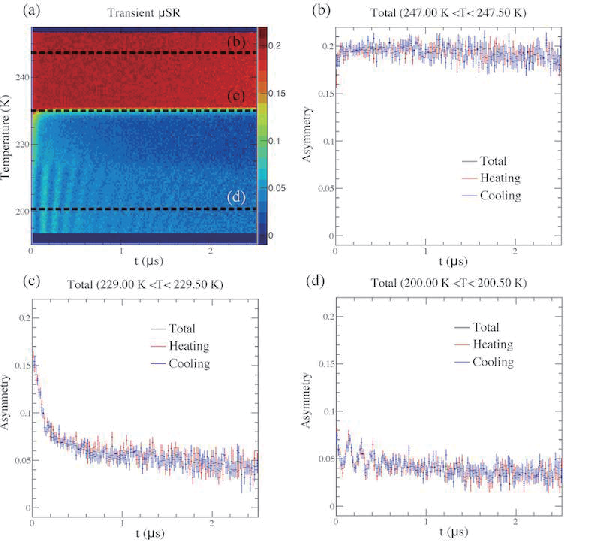}
 \caption{(a) Multi-dimensional histogram [$\tilde{A}(t)$] generated by event-analysis software,
showing the temperature and time on the vertical and horizontal axes, respectively,
 and the colors representing the magnitude of the asymmetry.
 (b), (c), and (d) are snapshots showing projections of the black line in (a) on the horizontal axis;
 these plots can be created by just moving the cursor.
}
 \label{Fig2}
\end{figure}

The event-analysis software developed for the transient $\mu$SR is based on ROOT (version 6.15/01 or later)~\cite{BRUN199781}, and requires an operating system with the ability to run bash shell scripts.   The analysis is performed in three stages, i.e., the data integration, histogram generation, and drawing parts. For the sake of speed, a simple confirmation-drawing mode is provided before creating the histogram. To provide an idea for the capability of the event-analysis software, snapshots of histogram manipulation using the graphic user interface (GUI) are shown in Fig.~\ref{Fig2} for the CuO data.  In the drawing part, by moving the cursor in the GUI, the time-dependent asymmetry can be displayed by carving the temperature or magnetic field into slices of arbitrary bin size.  In addition, the time-integrated asymmetry can be plotted as a function of the temperature or magnetic field [see Fig.~\ref{Fig3}(a)].  These features are useful in understanding the global behavior of a large number of spectra.

The ultimate goal of the $\mu$SR experiment is to extract information on physical properties of target materials (magnetism, superconductivity, etc.) based on a physical model from $\tilde{A}(t)$ [$\tilde{G}(t;T,B,...)$], and to distinguish it from the event analysis, such analysis will be called ``physical analysis'' in the following.
Although it is attractive to perform physical analysis on $\tilde{A}(t)$ by ``manifold (multi-dimensional surface) fitting'' directly on ROOT, it is yet to be implemented. Therefore, we will show two examples of analyzing $\tilde{A}(t)$ sliced for a sample condition of interest, using the batch processing function (chain fitting) of existing curve fitting programs.  For the data slicing, the event-analysis software is equipped with a one-click function to export spectra in the data format handled by musrfit~\cite{SUTER201269} or WiMDA~\cite{PRATT2000710}.  Since the data slicing is such that each sample condition parameter (e.g., temperature, magnetic field, etc.)  is orthogonal to each other, the potential complexity of error propagation is reduced to that within each histogram via the least-squares fits. The error propagation associated with direct analysis of multidimensional histograms, including the case of slicing in which sample condition parameter varies in a mixed manner, is a future issue to be incorporated with the development of the physical models with multidimensional parameters.

\subsection{$\mu$SR Experiment}
\noindent
The general-purpose $\mu$SR spectrometer, ARTEMIS, installed at the S1 area in MLF, J-PARC was used for the development of transient $\mu$SR technique; it accepts a surface muon beam with a total flux of $4.6\times10^5$~muon/s (double pulse mode) at the current proton beam operation of 0.8~MW and a repetition rate of 25~Hz~\cite{qubs1010011}.  ARTEMIS has two sets of positron hodoscopes consisting of 20 module detectors, KALLIOPE, arranged along the forward and backward positions relative to the sample; each KALLIOPE module harnesses 32 plastic scintillators ($12\ {\rm mm}\times 10\ {\rm mm}\times 10\ {\rm mm} $) with silicon photomultipliers, making up 16 ``telescopes'' (pairs of scintillators to identify the position trajectory), and signals are processed via on-chip readout circuits and Ethernet interfaces (SiTCP), through which data from 1280 scintillators in total are transferred to the data-acquisition (DAQ) PC~\cite{kojima_2014}. 

For the test measurements, CuO (99.99\% powder, supplied by Furuuchi Chemical Co.) and high-purity copper (99.9999\% plate, $50\times50\times1$ mm$^3$,  supplied by Nilaco Co.)  were adopted for temperature and field sweep, respectively.  CuO powder was put into a slab with $25\times25\times\sim$1 mm$^3$ on a silver plate.  The samples were loaded to a He gas-flow cryostat furnished with the ``fly-past'' vacuum vessel to eliminate backgrounds from muons stopped outside the sample.  ARTEMIS is also equipped with normal-conducting magnets, which can apply longitudinal and transverse magnetic fields of up to 0.4~T and 12.5~mT, respectively.

\section{Examples of Transient $\mu$SR}
\subsection{Temperature sweep: CuO}
\begin{figure}[h]
 \centering
 \includegraphics[width=0.38\textwidth,clip]{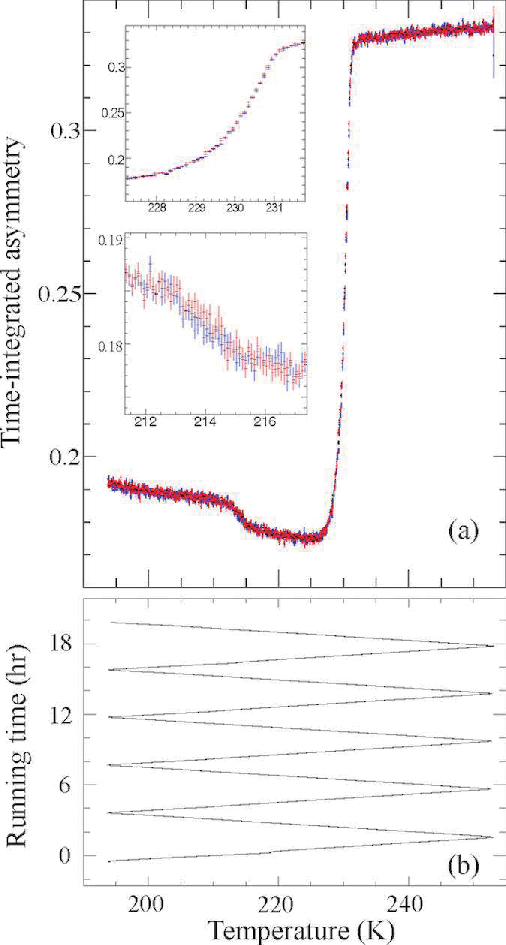}
 \caption{(a) Time-integrated asymmetry (red, blue, and black dots correspond to heating, cooling, and their mean, respectively), (b) real-time sample temperature monitored during the transient $\mu$SR measurement.  Insets: (a) the enlarged plots from around 227~K to 232~K and from 211~K to 217~K.
}
 \label{Fig3}
\end{figure}

\noindent
CuO is an antiferromagnet showing two magnetic transitions, a second-order transition at $T_{N1}\simeq230$~K and a first-order transition at $T_{N2}\simeq213$--215~K. In the intermediate temperature region between $T_{N1}$--$T_{N2}$, CuO is considered to exhibit an incommensurate magnetic ordered phase \cite{Rebello:13,Yang:89,Nishiyama2001}. Therefore, we tested how accurately we could capture these successive phase transitions with our new method. The temperature sweep range was 193~K to 250~K, the sweep rate was 0.5~K/min, and the proton beam power was about 0.51~MW at the time of measurement. A record of the temperature change during the measurement is shown in Fig.~\ref{Fig3}(b). The spontaneous muon precession frequency due to the internal magnetic field changes significantly on passing through  $T_{N2}$ from commensurate to incommensurate magnetic ordered phases with increasing temperature, where the actual time-differential $\mu$SR spectrum changes from partial oscillation ($\propto\frac{2}{3}\cos\omega_\mu t+\frac{1}{3}$, 
whose time-integral is $\frac{1}{3}$) to that approximated by the Gaussian Kubo-Toyabe relaxation with relatively large fluctuation rate (which leads to a reduction of the 1/3 component, see below).
Therefore, we can observe the reduction of integrated-asymmetry for $T>T_{N2}$ as shown in Fig.~\ref{Fig3}(a). 

In general, if the heat conduction between the sample and the sample holder is not sufficient, there will be a difference between the temperature indicated by the thermometer on the holder and the actual sample temperature. In this measurement, the fact that the sample was a powder may have also made temperature control by thermal conduction difficult. However, even in such a case, the temperature difference could be made sufficiently small by slowing down the temperature change sufficiently. This was confirmed by the fact that there was no difference in the time spectrum between heating and cooling, as shown in Fig.~\ref{Fig2}(b)--(d).

For effective physical analysis, it is useful to know in advance the general trend of the time spectra from the event-analysis results. In experiments with pulsed beams, the spin rotation signal is not observed above $f_{\rm N}$, which is determined by $1/2\delta$ with respect to the pulse width $\delta$. The presence of such an internal magnetic field appears as a reduction of the initial asymmetry. (In the case of MUSE, $\delta$ is FWHM $\simeq$80~ns and  $f_{\rm N}\simeq6.25$~MHz.) Furthermore, when integrated over time, the contribution of all spin rotation signals becomes zero, which is observed as a decrease in the integral asymmetry. Figure \ref{Fig3}(a) shows the temperature dependence of the asymmetry with integration performed for each temperature change in heating and cooling. Although the expected hysteresis gap of $\sim$2~K near $T_{N2}$ is much greater than the temperature step, it was not observed. As suggested by this result, it should be noted that the integral method, which cancels out the contribution of the rotational signal, may not provide sufficient sensitivity to the magnetic hysteresis.

\begin{figure}[t]
 \centering
 \includegraphics[width=0.4\textwidth,clip]{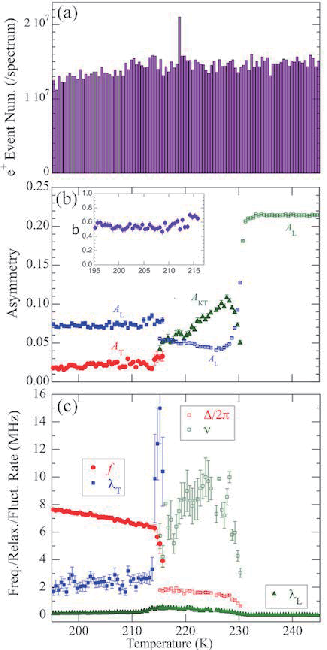}
 \caption{(a) Positron event statistics per sliced $\mu$SR spectrum, (b) partial asymmetry, and (c) frequency/relaxation/fluctuation rates vs.~temperature.  Insets: (b) the power of the stretched exponential decay in Eq.(\ref{strtch}).
}
 \label{Fig4}
\end{figure}

As is clear from the behavior of $\tilde{A}(t)$ in Fig.~\ref{Fig2}, the time spectra are roughly described as a slow relaxation in the paramagnetic (PM) phase above $T_{\rm N1}$, the sum of two different relaxation components in the incommensurate antiferromagnetic (IAFM) phase of $T_{\rm N1}$--$T_{\rm N2}$, and the sum of a rotation component and a relaxation component in the antiferromagnetic (AFM) phase below $T_{\rm N2}$. Therefore, we generated 120 $\mu$SR time spectra [$A(t)$] by slicing $\tilde{A}(t)$ with a temperature bin of 0.5 K, and analyzed them by musrfit. As shown in Fig.~\ref{Fig4}(a), the statistics (sum of those obtained during temperature rise and fall) for each spectrum was approximately 12-14 M events. Curve fitting was performed assuming the following equations for the respective temperature regions,
\begin{equation}
A(t) = \left\{
\begin{array}{ll}
A_{\rm L}e^{-\lambda_{\rm L} t}, & ({\rm PM})\\
A_{\rm KT}G_{\rm KT}(t;\Delta,\nu)+A_{\rm L}e^{-(\lambda_{\rm L} t)^\beta}, & ({\rm IAFM})\\
A_{\rm T}e^{-\lambda_{\rm T} t}\cos\omega_\mu t +A_{\rm L}e^{-\lambda_{\rm L} t}, &({\rm AFM})
\end{array} \label{strtch}
\right.
\end{equation}
where $\lambda_{\rm L}$ ($\lambda_{\rm T}$) is the longitudinal (transverse) relaxation rate with $A_{\rm L}$ ($A_{\rm T}$) being the partial asymmetry, $\beta$ is the power for the stretched exponential damping, $\omega_\mu \equiv 2\pi f=\gamma_\mu B_{\rm loc}$ is the muon Larmor frequency with $\gamma_\mu=2\pi\times135.53$~MHz/T being the muon gyromagnetic ratio and $B_{\rm loc}$ the internal field at the muon site, $G_{\rm KT}(t;\Delta,\nu)$ is the Gaussian Kubo-Toyabe function with $A_{\rm KT}$ the partial asymmetry, which approximates $B_{\rm loc}$ due to incommensurate magnetic ordering with a random magnetic field by its rms value $\langle B_{\rm loc}^2\rangle^{1/2}=\Delta/\gamma_\mu$ with the fluctuation rate $\nu$ \cite{Hayano:79}. Prior to this analysis, the instrumental asymmetry $\alpha$ was determined by a curve fit that takes into account the fact that $A(t)$ appears to converge to zero at late times in the IAFM phase, and was fixed to that value for all time spectra in the subsequent analysis. The analysis of all 120 spectra by sequential fitting took about 10 minutes, allowing repeated analysis by trial-and-error of the fitting functions. The temperature dependence of each parameter obtained by curve fitting is shown in Fig.~\ref{Fig4}b) and (c) for the 195--245~K range.

As is found in Fig.~\ref{Fig4}(b), the value of $A_{\rm L}$ in the PM phase is about 0.21, which corresponds to the total asymmetry [$A_0$ in Eq.~(\ref{asy})], while in the AFM phase it is 0.07, about 1/3 of that value. This indicates that the entire sample volume is in an ordered state; one would expect 
$$G(t)= \frac{1}{3}e^{-\lambda_{\rm L} t}+\frac{2}{3}\sum_i y_i\exp(-\lambda_{\rm T}^{(i)} t)\cos\omega_\mu^{(i)}$$ 
as a powder average, where $\omega_\mu^{(i)}$ and $\lambda_{\rm T}^{(i)}$ denote the rotation frequency and relaxation rate at the $i$th muon site having relative yield $y_i$ ($\sum_i y_i=1$)  \cite{Yaouanc:00,Blundell:21}.  Meanwhile, the amplitude $A_{\rm T}$ of the rotational component in the AFM phase is much smaller than the expected value of $0.21\times2/3 = 0.14$ in such a case. This can be qualitatively explained by the fact that the observed rotation frequency $f$ [Fig.~\ref{Fig4}(c)] is higher than the Nyquist frequency. Previous studies with DC muon beams have reported that muons are distributed over several magnetically non-equivalent sites that exhibit greater $B_{\rm loc}$ \cite{Nishiyama2001}, and the disappearance of the corresponding high-frequency rotation signals also contributes to the decrease in $A_{\rm T}$.

In contrast to the AFM phase, no clear rotational signal is observed in the IAFM phase, and the time spectra are reproduced by a component showing Gaussian-type fast relaxation and a longitudinal relaxation component. The static linewidth $\Delta$ obtained by fitting with the Kubo-Toyabe function is about 10--12~MHz near $T_{\rm N2}$, and when this Gaussian damping is interpreted to be the first quarter period of the rotation signal, the corresponding frequency is $\Delta/2\pi\sim1.6$--2~MHz, which is smoothly extrapolated from $f$ just below $T_{\rm N2}$. Therefore, it is suggested that $\Delta$ represents the static distribution width determined by the configuration of Cu magnetic moments, and its fluctuation frequency $\nu$ corresponds to that of the Cu moment fluctuation. In other words, we can conclude that magnetic fluctuations of 4--8~MHz exist in the IAFM phase. This is also supported by the behavior of the LF spectra taken in parallel with ZF data (not shown). 

\subsection{Field sweep: Cu}
Resonance spectroscopy is a powerful method because it can focus on any specific part of the interaction in matter. One such technique in $\mu$SR is well known as $\mu$LCR, similar to the level crossing (or ``cross relaxation'') method commonly used in solid-state NMR. In non-magnetic metals, $\mu$LCR occurs between the Zeeman levels of muon spin ${\bm S}$ under an external magnetic field ${\bm B}$ and the electric quadrupole levels of the adjacent host nuclei (spin $I\ge1$). The local symmetry of the surrounding host lattice is broken by the electron-lattice distortion caused by the presence of $\mu^+$, and electric quadrupole interaction is induced between nuclear quadrupoles and electric field gradient (EFG), partly removing the degeneracy of nuclear spin states:
\begin{equation}
\omega_I(I_z) =\frac{1}{2}\omega_{\rm Q}\left[I_z^2-\frac{1}{3}I(I+1)+u(I_z,\eta^2)\right],\\
\end{equation}
\begin{equation}
\hspace{2em}\omega_{\rm Q}=\frac{3e^2qQ}{2\hbar I(2I-1)},
\end{equation}
where $eq\equiv V_{zz}$ is the EFG at the nuclear sites, $eQ$ is the nuclear quadrupole moment, and $\eta$ is the asymmetry of EFG, $\eta=(V_{xx}-V_{yy})/V_{zz}$ (with $|V_{zz}|\ge |V_{xx}| \ge |V_{yy}|$). The last term $u$ is of important relevance to $\mu$LCR in that the symmetry of EFG may be affected by the presence of $\mu^+$.
In copper crystals, for example, the nuclear spin levels ($I=3/2$) are split with an energy $\omega_I(\pm\frac{3}{2})-\omega_I(\pm\frac{1}{2})=\omega_{\rm Q}$.

In the conventional $\mu$LCR experiment, the resonance is detected as an enhancement of longitudinal spin relaxation, for which muons polarized parallel to ${\bm B}$ are used. The magnitude of ${\bm B}$ is varied to match the energy splitting of the muon's Zeeman level with that of the magnetic level of the host nucleus determined primarily by the electric quadrupole energy.
Usually, the magnetic dipole interaction (described by the Hamiltonian $H_{\rm D}$) between muon and nucleus is much smaller than the Zeeman interaction of muons. In this case, the spin relaxation can be approximated by the $n$th-order parallel moment $M_n'$ corresponding to the diagonal term of $H_{\rm D}$  as follows \cite{Kreitzman:86a,Kreitzman:86b}:
\begin{eqnarray}
G_z(t)&\simeq&1-\frac{1}{2}M_2't^2+\frac{1}{4!}M_4't^4-... ,\\
M_4'(B)&=&M_2'\omega_D^2 \sum_{i=1}^{N}[(\omega_{\rm Q}-\gamma_\mu B)^2\nonumber\\ 
& & +\: C(\Theta_i)(\omega_{\rm Q}-\gamma_\mu B)\gamma_{\rm Cu}B \nonumber\\ 
& & +\: D(\Theta_i,I_z)\gamma_{\rm Cu}^2B^2],
\end{eqnarray}
where $\gamma_{\rm Cu}$ is the gyromagnetic ratio of Cu nucleus ($=2\pi\times11.285$ and 12.089~MHz/T for $^{63}$Cu and $^{65}$Cu, respectively), $\omega_D=\gamma_\mu\gamma_{\rm Cu}/r^3$ ($r$ being the distance between muon and Cu nucleus),  $C$ [$=(1-3\sin^2\Theta_i)/(1+\frac{3}{2}\sin^2\Theta_i)$]
and $D$ are geometrical factors determined by the polar angle $\Theta_i$ between the EFG and the muon-nuclear axis for the $i$th nucleus, and $N$ is the number of the nearest neighboring Cu nuclei.  Since $M'_2$ is a constant, the $\mu$LCR condition
is then given as a field $B_{\rm res}$ for minimizing $M_4'$ (note that the sign of the coefficient for $M_4'$ is positive), i.e.,
\begin{equation}
B_{\rm res} \simeq \frac{\omega_{\rm Q}}{\gamma_\mu}\left[1 + \frac{\overline{C}\gamma_{\rm Cu}}{2\gamma_\mu}+O\left(\frac{\gamma_{\rm Cu}^2}{\gamma_\mu^2}\right)\right],
\end{equation}
where $\overline{C}=\sum_iC(\Theta_i)/N$.  Since $\overline{C}\le 1$ and $D\sim1$, the above equation is approximated to yield $B_{\rm res}\simeq\omega_{\rm Q}/\gamma_\mu$. 

\begin{figure}[]
 \centering
 \includegraphics[width=0.48\textwidth,clip]{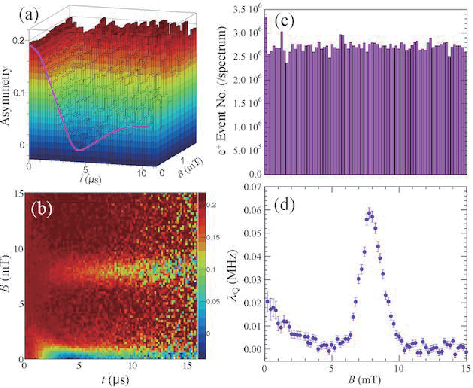}
 \caption{Results of the LCR measurements on a copper plate.
(a) A 3D plot of $\tilde{A}(t)$ with the longitudinal magnetic field $B$ for the direction crossing the screen surface. The solid pink line represents the static Kubo-Toyabe function observed under zero field ($B=0$). (b) A contour plot of $\tilde{A}(t)$ at 50~K, in which the vertical and horizontal axes represent $B$ and $t$, respectively. (c) Positron event statistics per sliced $\mu$SR spectrum, and (d) relaxation rate vs. temperature.}
 \label{Fig5}
\end{figure}

The $\mu$LCR measurements on copper were performed over a magnetic field range covering $B_{\rm res}\simeq 8$~mT at a few different temperatures.  An example of relatively high-statistics $\tilde{A}(t)$ measured at 50~K is shown in Fig.~\ref{Fig5}(a) and (b), where the field was swept from 0 to 15.0~mT in 0.2~mT steps ($\sim$80 s/step, with the elapsed time of 1.7 hr) under a proton beam power of $\sim$0.60 MW.  The Kubo-Toyabe function near $B=0$ and the suppression of relaxation by increasing $B$ can be observed in the contour plot. The number of positron events per spectrum sliced at 0.2~mT (the same as the $B$ step) is shown in Fig.~\ref{Fig5}(c). The physical analysis for 76 spectra were performed using musrfit, where the curve fits were made (with sequential fitting) to the following form,
\begin{equation}
A(t)=A_0G_{\rm KT}(t;\Delta_{\rm n},B)e^{-\lambda_{\rm Q} t},
\end{equation}
and $G_{\rm KT}(t;\Delta_{\rm n},B)$ is the static Kubo-Toyabe function under a longitudinal field $B$ with $\Delta_{\rm n}$ [$\simeq(M_2'/2)^{1/2}$] being the linewidth determined by the random local fields from Cu nuclear magnetic moments. The curve fits of the spectra for $B\le 1.0$~mT yielded $\Delta_{\rm n}=0.374(9)$~MHz in excellent agreement with the earlier results \cite{Kadono:89,Luke:91}.  (The muon hopping frequency, which determines the fluctuation rate $\nu_{\rm n}$ of $\Delta_{\rm n}$, is known to be negligibly small at 50~K in copper \cite{Kadono:89,Luke:91}, and assumed to be zero.) The obtained $\lambda_{\rm Q}$ versus $B$ is shown in Fig.~\ref{Fig5}(d).  The peak of $\lambda_{\rm Q}$ is observed at 7.78~mT, which is in excellent agreement with the previous experimental result~\cite{Kreitzman:86a,Kreitzman:86b}.
Since we used a standard DC power supply, continuous sweep of the magnetic field was not possible. 
However, the magnetic field was changed asynchronously with the DAQ, which was a good demonstration of continuous and asynchronous magnetic field sweep like the magnetic field of a superconducting magnet in the future.

\section{Summary and Prospects}
\noindent
In this paper, we establish a transient $\mu$SR method consisting of (i) data acquisition in pulse-by-pulse mode that integrates information on the sample environment in real time and (ii) event-analysis software that generates a multidimensional asymmetry histogram from the obtained data. The transient $\mu$SR method was applied to antiferromagnetic CuO and metallic Cu samples to demonstrate its ability to measure transient time spectra for varying temperature and magnetic field in each case. Transient $\mu$SR, in combination with a high-flux pulsed muon beam, enables fast scanning of the sample environment, which is difficult with conventional measurements, and makes it possible to measure time-dependent phenomena in materials, such as operando measurements. The method can also be applied to other pulsed muon beam experiments such as muon X-ray measurements. 

The time scale of change in sample conditions over which this technique is valid is so far limited by the interval between beam pulses (40 ms). This is partly due to the relatively long time constant of the sample environment under normal circumstances. However, it is expected that this limitation can be overcome by combining this technique with a stroboscopic measurement method that synchronizes the timing of the beam pulses with the fast changing (or pulsed) conditions \cite{Nojiri:11,Nakajima:18}. This will be a future challenge in further developing transient $\mu$SR.

Another major challenge is to perform direct physical analysis instead of slicing multidimensional histograms and analyzing them with conventional physical analysis software. 
In this regard, we are focusing on the ``deep learning'', 
which has been rapidly developing in recent years. Deep learning is revolutionary in that it has shown the possibility of automating the act of constructing physical models; 
AI-Feynman, for example, has attempted to derive physical laws with some success \cite{Udrescu:20}. 
For $\mu$SR, an attempt has been made to analyze the time spectra using a simple machine learning code called Principal Component Analysis (PCA) \cite{Tula_2022}.  While this showed the possibility of parametrizing some features of the data without relying on a specific physical model, there still seems to be significant difficulties in bridging these abstract parameters to a physical model comprehensible to humans.  The development of analysis software that learns the direct correspondence between multidimensional histograms and physical models through deep learning would be a great step forward in solving this problem.

\section{Acknowledgements}
 The authors would like to acknowledge the help and expertise of the MUSE S-line staff during experiment at MLF, J-PARC.
 The muon experiment was conducted under the support of Inter-University-Research Programs by Institute of Materials Structure Science, KEK (Proposals No.~2019B0411 and 2020MI21). This work was partially supported by JSPS KAKENHI (Grant No. 20K05312, 19K15033, and 19K20595) and the MEXT Elements Strategy Initiative to Form Core Research Center for Electron Materials (Grant No. JPMXP0112101001).



  \bibliographystyle{elsarticle-num} 
  \bibliography{ref}

\begin{thebibliography}{10}
\expandafter\ifx\csname url\endcsname\relax
  \def\url#1{\texttt{#1}}\fi
\expandafter\ifx\csname urlprefix\endcsname\relax\def\urlprefix{URL }\fi
\expandafter\ifx\csname href\endcsname\relax
  \def\href#1#2{#2} \def\path#1{#1}\fi

\bibitem{Klempt:82}
E.~Klempt, R.~Schulze, H.~Wolf, M.~Camani, F.~N. Gygax, W.~R\"uegg, A.~Schenck,
  H.~Schilling,
  \href{https://link.aps.org/doi/10.1103/PhysRevD.25.652}{Measurement of the
  magnetic moment of the positive muon by a stroboscopic muon-spin-rotation
  technique}, Phys. Rev. D 25 (1982) 652--676.
\newblock \href {https://doi.org/10.1103/PhysRevD.25.652}
  {\path{doi:10.1103/PhysRevD.25.652}}.
\newline\urlprefix\url{https://link.aps.org/doi/10.1103/PhysRevD.25.652}

\bibitem{Schenck:85}
A.~Schenck, Muon spin rotation spectroscopy: principles and applications in
  solid state physics, Taylor and Francis Inc., Philadelphia, PA, 1985.

\bibitem{Ishida:84}
K.~Ishida, T.~Matsuzaki, K.~Nishiyama, K.~Nagamine,
  \href{https://doi.org/10.1007/BF02066138}{Time dependent local field felt by
  $\mu^+$ in {MnO} revealed by spin resonance method}, Hyperfine Interact.
  19~(1) (1984) 927--931.
\newblock \href {https://doi.org/10.1007/BF02066138}
  {\path{doi:10.1007/BF02066138}}.
\newline\urlprefix\url{https://doi.org/10.1007/BF02066138}

\bibitem{Nishiyama:86}
K.~Nishiyama, T.~Azuma, K.~Ishida, T.~Matsuzaki, J.~Imazato, T.~Yamazaki,
  K.~Nagamine, \href{https://doi.org/10.1007/BF02394999}{Development of 500
  {MHz} muon spin resonance spectrometer}, Hyperfine Interact. 32~(1) (1986)
  887--892.
\newblock \href {https://doi.org/10.1007/BF02394999}
  {\path{doi:10.1007/BF02394999}}.
\newline\urlprefix\url{https://doi.org/10.1007/BF02394999}

\bibitem{Kadono:94}
R.~Kadono, A.~Matsushita, R.~M. Macrae, K.~Nishiyama, K.~Nagamine,
  \href{https://link.aps.org/doi/10.1103/PhysRevLett.73.2724}{Muonium centers
  in crystalline {Si} and {Ge} under illumination}, Phys. Rev. Lett. 73 (1994)
  2724--2727.
\newblock \href {https://doi.org/10.1103/PhysRevLett.73.2724}
  {\path{doi:10.1103/PhysRevLett.73.2724}}.
\newline\urlprefix\url{https://link.aps.org/doi/10.1103/PhysRevLett.73.2724}

\bibitem{Kadono:03}
R.~Kadono, R.~M. Macrae, K.~Nagamine,
  \href{https://link.aps.org/doi/10.1103/PhysRevB.68.245204}{Charge dynamics of
  muonium centers in {Si} revealed by photoinduced muon spin relaxation}, Phys.
  Rev. B 68 (2003) 245204.
\newblock \href {https://doi.org/10.1103/PhysRevB.68.245204}
  {\path{doi:10.1103/PhysRevB.68.245204}}.
\newline\urlprefix\url{https://link.aps.org/doi/10.1103/PhysRevB.68.245204}

\bibitem{Ghandi:07}
K.~Ghandi, I.~P. Clark, J.~S. Lord, S.~P. Cottrell,
  \href{http://dx.doi.org/10.1039/B615184C}{Laser-muon spin spectroscopy in
  liquids---a technique to study the excited state chemistry of transients},
  Phys. Chem. Chem. Phys. 9 (2007) 353--359.
\newblock \href {https://doi.org/10.1039/B615184C}
  {\path{doi:10.1039/B615184C}}.
\newline\urlprefix\url{http://dx.doi.org/10.1039/B615184C}

\bibitem{Shimomura:12}
K.~Shimomura, P.~Bakule, F.~Pratt, K.~Ishida, K.~Ohishi, I.~Watanabe,
  Y.~Matsuda, K.~Nagamine, E.~Torikai, K.~Nishiyama,
  \href{https://www.sciencedirect.com/science/article/pii/S1875389212012643}{Photo
  detachment of negatively charged muonium in {GaAs} by laser irradiation},
  Physics Procedia 30 (2012) 224--226.
\newblock \href {https://doi.org/https://doi.org/10.1016/j.phpro.2012.04.078}
  {\path{doi:https://doi.org/10.1016/j.phpro.2012.04.078}}.
\newline\urlprefix\url{https://www.sciencedirect.com/science/article/pii/S1875389212012643}

\bibitem{Yokoyama:17}
K.~Yokoyama, J.~S. Lord, J.~Miao, P.~Murahari, A.~J. Drew,
  \href{https://link.aps.org/doi/10.1103/PhysRevLett.119.226601}{Photoexcited
  muon spin spectroscopy: A new method for measuring excess carrier lifetime in
  bulk silicon}, Phys. Rev. Lett. 119 (2017) 226601.
\newblock \href {https://doi.org/10.1103/PhysRevLett.119.226601}
  {\path{doi:10.1103/PhysRevLett.119.226601}}.
\newline\urlprefix\url{https://link.aps.org/doi/10.1103/PhysRevLett.119.226601}

\bibitem{Yokoyama:21}
K.~Yokoyama, J.~S. Lord, J.~Miao, P.~Murahari, A.~J. Drew,
  \href{https://doi.org/10.1063/5.0054291}{{Decoupling bulk and surface
  recombination properties in silicon by depth-dependent carrier lifetime
  measurements}}, Appl. Phys. Lett. 118~(25) (2021) 252105.
\newblock \href
  {http://arxiv.org/abs/https://pubs.aip.org/aip/apl/article-pdf/doi/10.1063/5.0054291/14549720/252105\_1\_online.pdf}
  {\path{arXiv:https://pubs.aip.org/aip/apl/article-pdf/doi/10.1063/5.0054291/14549720/252105\_1\_online.pdf}},
  \href {https://doi.org/10.1063/5.0054291} {\path{doi:10.1063/5.0054291}}.
\newline\urlprefix\url{https://doi.org/10.1063/5.0054291}

\bibitem{Murphy:22}
J.~D. Murphy, N.~E. Grant, S.~L. Pain, T.~Niewelt, A.~Wratten, E.~Khorani,
  V.~P. Markevich, A.~R. Peaker, P.~P. Altermatt, J.~S. Lord, K.~Yokoyama,
  \href{https://doi.org/10.1063/5.0099492}{{Carrier lifetimes in high-lifetime
  silicon wafers and solar cells measured by photoexcited muon spin
  spectroscopy}}, J. Appl. Phys. 132~(6) (2022) 065704.
\newblock \href
  {http://arxiv.org/abs/https://pubs.aip.org/aip/jap/article-pdf/doi/10.1063/5.0099492/16514408/065704\_1\_online.pdf}
  {\path{arXiv:https://pubs.aip.org/aip/jap/article-pdf/doi/10.1063/5.0099492/16514408/065704\_1\_online.pdf}},
  \href {https://doi.org/10.1063/5.0099492} {\path{doi:10.1063/5.0099492}}.
\newline\urlprefix\url{https://doi.org/10.1063/5.0099492}

\bibitem{Motokawa:91}
M.~Motokawa, H.~Nojiri, M.~Uchi, S.~Watamura, K.~Nishiyama, K.~Nagamine,
  \href{https://doi.org/10.1007/BF02397765}{Application of pulsed high magnetic
  field to {$\mu$SR} studies}, Hyperfine Interact. 65~(1) (1991) 1089--1095.
\newblock \href {https://doi.org/10.1007/BF02397765}
  {\path{doi:10.1007/BF02397765}}.
\newline\urlprefix\url{https://doi.org/10.1007/BF02397765}

\bibitem{Shiroka:99}
T.~Shiroka, C.~Bucci, R.~De~Renzi, F.~Galli, G.~Guidi, G.~H. Eaton, P.~J.~C.
  King, C.~A. Scott,
  \href{https://link.aps.org/doi/10.1103/PhysRevLett.83.4405}{Polarized muon
  spins in pulsed magnetic fields: A new method to study delayed muonium
  formation}, Phys. Rev. Lett. 83 (1999) 4405--4408.
\newblock \href {https://doi.org/10.1103/PhysRevLett.83.4405}
  {\path{doi:10.1103/PhysRevLett.83.4405}}.
\newline\urlprefix\url{https://link.aps.org/doi/10.1103/PhysRevLett.83.4405}

\bibitem{Cottrell:12}
S.~Cottrell, F.~Pratt, A.~Hillier, P.~King, F.~Akeroyd, A.~J. Markvardsen,
  N.~Draper, Y.~Yao, S.~Blundell,
  \href{https://www.sciencedirect.com/science/article/pii/S1875389212012175}{Data
  formats and analysis codes--new software for {$\mu$SR}}, Phys. Procedia 30
  (2012) 20--25.
\newblock \href {https://doi.org/https://doi.org/10.1016/j.phpro.2012.04.031}
  {\path{doi:https://doi.org/10.1016/j.phpro.2012.04.031}}.
\newline\urlprefix\url{https://www.sciencedirect.com/science/article/pii/S1875389212012175}

\bibitem{Giblin:14}
S.~Giblin, S.~Cottrell, P.~King, S.~Tomlinson, S.~Jago, L.~Randall, M.~Roberts,
  J.~Norris, S.~Howarth, Q.~Mutamba, N.~Rhodes, F.~Akeroyd,
  \href{https://www.sciencedirect.com/science/article/pii/S016890021400285X}{Optimising
  a muon spectrometer for measurements at the {ISIS} pulsed muon source}, Nucl.
  Instr. Meth. Phys. Res. Sect. A: Accelerators, Spectrometers, Detectors and
  Associated Equipment 751 (2014) 70--78.
\newblock \href {https://doi.org/https://doi.org/10.1016/j.nima.2014.03.010}
  {\path{doi:https://doi.org/10.1016/j.nima.2014.03.010}}.
\newline\urlprefix\url{https://www.sciencedirect.com/science/article/pii/S016890021400285X}

\bibitem{PETERSON201524}
P.~F. Peterson, S.~I. Campbell, M.~A. Reuter, R.~J. Taylor, J.~Zikovsky,
  \href{https://www.sciencedirect.com/science/article/pii/S0168900215010682}{Event-based
  processing of neutron scattering data}, Nuclear Instruments and Methods in
  Physics Research Section A: Accelerators, Spectrometers, Detectors and
  Associated Equipment 803 (2015) 24--28.
\newblock \href {https://doi.org/https://doi.org/10.1016/j.nima.2015.09.016}
  {\path{doi:https://doi.org/10.1016/j.nima.2015.09.016}}.
\newline\urlprefix\url{https://www.sciencedirect.com/science/article/pii/S0168900215010682}

\bibitem{ARNOLD2014156}
O.~Arnold, J.~Bilheux, J.~Borreguero, A.~Buts, S.~Campbell, L.~Chapon,
  M.~Doucet, N.~Draper, R.~{Ferraz Leal}, M.~Gigg, V.~Lynch, A.~Markvardsen,
  D.~Mikkelson, R.~Mikkelson, R.~Miller, K.~Palmen, P.~Parker, G.~Passos,
  T.~Perring, P.~Peterson, S.~Ren, M.~Reuter, A.~Savici, J.~Taylor, R.~Taylor,
  R.~Tolchenov, W.~Zhou, J.~Zikovsky,
  \href{https://www.sciencedirect.com/science/article/pii/S0168900214008729}{Mantid--data
  analysis and visualization package for neutron scattering and $\mu$sr
  experiments}, Nuclear Instruments and Methods in Physics Research Section A:
  Accelerators, Spectrometers, Detectors and Associated Equipment 764 (2014)
  156--166.
\newblock \href {https://doi.org/https://doi.org/10.1016/j.nima.2014.07.029}
  {\path{doi:https://doi.org/10.1016/j.nima.2014.07.029}}.
\newline\urlprefix\url{https://www.sciencedirect.com/science/article/pii/S0168900214008729}

\bibitem{Yaouanc:00}
A.~Yaouanc, P.~Dalmas~de R\'eotier, Muon Spin Rotation, Relaxation and
  Resonance: Applications to Condensed Matter, International Series of
  Monographs on Physics, Oxford University Press, Oxford, 2010.

\bibitem{Blundell:21}
S.~J. Blundell, R.~De~Renzi, T.~Lancaster, F.~L. Pratt, Muon Spectroscopy: An
  Introduction, Oxford University Press, Oxford, 2021.

\bibitem{BRUN199781}
R.~Brun, F.~Rademakers, Root - an object oriented data analysis framework, in:
  AIHENP'96 Workshop, Lausane, Vol. 389, 1996, pp. 81--86.

\bibitem{SUTER201269}
A.~Suter, B.~M. Wojek, Musrfit: {A} free platform-independent framework for
  $\mu${SR} data analysis, Physics Procedia 30 (2012) 69.
\newblock \href {https://doi.org/https://doi.org/10.1016/j.phpro.2012.04.042}
  {\path{doi:https://doi.org/10.1016/j.phpro.2012.04.042}}.

\bibitem{PRATT2000710}
F.~L. Pratt, {WiMDA}: a muon data analysis program for the {Windows} {PC},
  Physica B: Condens. Matter 289-290 (2000) 710--714.
\newblock \href {https://doi.org/https://doi.org/10.1016/S0921-4526(00)00328-8}
  {\path{doi:https://doi.org/10.1016/S0921-4526(00)00328-8}}.

\bibitem{qubs1010011}
W.~Higemoto, R.~Kadono, N.~Kawamura, A.~Koda, K.~M. Kojima, S.~Makimura,
  S.~Matoba, Y.~Miyake, K.~Shimomura, P.~Strasser, {Materials and Life Science
  Experimental Facility at the Japan Proton Accelerator Research Complex IV:
  The Muon Facility}, Quantum Beam Science 1~(1) (2017).
\newblock \href {https://doi.org/10.3390/qubs1010011}
  {\path{doi:10.3390/qubs1010011}}.

\bibitem{kojima_2014}
K.~M. Kojima, T.~Murakami, Y.~Takahashi, H.~Lee, S.~Y. Suzuki, A.~Koda,
  I.~Yamauchi, M.~Miyazaki, M.~Hiraishi, H.~Okabe, S.~Takeshita, R.~Kadono,
  T.~Ito, W.~Higemoto, S.~Kanda, Y.~Fukao, N.~Saito, M.~Saito, M.~Ikeno,
  T.~Uchida, M.~M. Tanaka, New $\mu${SR} spectrometer at {J}-{PARC} {MUSE}
  based on {Kalliope} detectors, J. Phys.: Conf. Ser. 551 (2014) 012063.
\newblock \href {https://doi.org/10.1088/1742-6596/551/1/012063}
  {\path{doi:10.1088/1742-6596/551/1/012063}}.

\bibitem{Rebello:13}
A.~Rebello, Z.~C.~M. Winter, S.~Viall, J.~J. Neumeier, Multiple phase
  transitions in cuo observed with thermal expansion, Phys. Rev. B 88 (2013)
  094420.
\newblock \href {https://doi.org/10.1103/PhysRevB.88.094420}
  {\path{doi:10.1103/PhysRevB.88.094420}}.

\bibitem{Yang:89}
B.~X. Yang, T.~R. Thurston, J.~M. Tranquada, G.~Shirane, Magnetic neutron
  scattering study of single-crystal cupric oxide, Phys. Rev. B 39 (1989)
  4343--4349.
\newblock \href {https://doi.org/10.1103/PhysRevB.39.4343}
  {\path{doi:10.1103/PhysRevB.39.4343}}.

\bibitem{Nishiyama2001}
K.~Nishiyama, W.~Higemoto, K.~Shimomura, A.~Koda, G.~Maruta, S.~W. Nishiyama,
  X.~G. Zheng, Multiple phase transitions in cuo studied by $\mu${SR},
  Hyperfine Interact. 136~(3) (2001) 289--294.
\newblock \href {https://doi.org/10.1023/A:1020504620851}
  {\path{doi:10.1023/A:1020504620851}}.

\bibitem{Hayano:79}
R.~S. Hayano, Y.~J. Uemura, J.~Imazato, N.~Nishida, T.~Yamazaki, R.~Kubo,
  Zero-and low-field spin relaxation studied by positive muons, Phys. Rev. B 20
  (1979) 850--859.
\newblock \href {https://doi.org/10.1103/PhysRevB.20.850}
  {\path{doi:10.1103/PhysRevB.20.850}}.

\bibitem{Kreitzman:86a}
S.~R. Kreitzman, J.~H. Brewer, D.~R. Harshman, R.~Keitel, D.~L. Williams, K.~M.
  Crowe, E.~J. Ansaldo, Longitudinal-field $\mu^+$ spin relaxation via
  quadrupolar level-crossing resonance in {Cu} at 20 {K}, Phys. Rev. Lett. 56
  (1986) 181--184.
\newblock \href {https://doi.org/10.1103/PhysRevLett.56.181}
  {\path{doi:10.1103/PhysRevLett.56.181}}.

\bibitem{Kreitzman:86b}
S.~R. Kreitzman, Muon level crossing resonance in diamagnetic systems-general
  considerations, Hyperfine Interact. 31~(1) (1986) 13--28.
\newblock \href {https://doi.org/10.1007/BF02401534}
  {\path{doi:10.1007/BF02401534}}.

\bibitem{Kadono:89}
R.~Kadono, J.~Imazato, T.~Matsuzaki, K.~Nishiyama, K.~Nagamine, T.~Yamazaki,
  D.~Richter, J.-M. Welter, Quantum diffusion of positive muons in copper,
  Phys. Rev. B 39 (1989) 23--41.
\newblock \href {https://doi.org/10.1103/PhysRevB.39.23}
  {\path{doi:10.1103/PhysRevB.39.23}}.

\bibitem{Luke:91}
G.~M. Luke, J.~H. Brewer, S.~R. Kreitzman, D.~R. Noakes, M.~Celio, R.~Kadono,
  E.~J. Ansaldo, Muon diffusion and spin dynamics in copper, Phys. Rev. B 43
  (1991) 3284--3297.
\newblock \href {https://doi.org/10.1103/PhysRevB.43.3284}
  {\path{doi:10.1103/PhysRevB.43.3284}}.

\bibitem{Nojiri:11}
H.~Nojiri, S.~Yoshii, M.~Yasui, K.~Okada, M.~Matsuda, J.~S. Jung, T.~Kimura,
  L.~Santodonato, G.~E. Granroth, K.~A. Ross, J.~P. Carlo, B.~D. Gaulin,
  Neutron laue diffraction study on the magnetic phase diagram of multiferroic
  ${\mathrm{mnwo}}_{4}$ under pulsed high magnetic fields, Phys. Rev. Lett. 106
  (2011) 237202.
\newblock \href {https://doi.org/10.1103/PhysRevLett.106.237202}
  {\path{doi:10.1103/PhysRevLett.106.237202}}.

\bibitem{Nakajima:18}
T.~Nakajima, Y.~Inamura, T.~Ito, K.~Ohishi, H.~Oike, F.~Kagawa, A.~Kikkawa,
  Y.~Taguchi, K.~Kakurai, Y.~Tokura, T.-h. Arima, Phase-transition kinetics of
  magnetic skyrmions investigated by stroboscopic small-angle neutron
  scattering, Phys. Rev. B 98 (2018) 014424.
\newblock \href {https://doi.org/10.1103/PhysRevB.98.014424}
  {\path{doi:10.1103/PhysRevB.98.014424}}.

\bibitem{Udrescu:20}
S.~M. Udrescu, M.~Tegmark, Ai feynman: A physics-inspired method for symbolic
  regression, Science Advances 6~(16) (2020) eaay2631.
\newblock \href {https://doi.org/10.1126/sciadv.aay2631}
  {\path{doi:10.1126/sciadv.aay2631}}.

\bibitem{Tula_2022}
T.~Tula, G.~M\"oller, J.~Quintanilla, S.~R. Giblin, A.~D. Hillier, E.~E.
  McCabe, S.~Ramos, D.~S. Barker, S.~Gibson,
  \href{https://dx.doi.org/10.1088/1742-6596/2164/1/012018}{Joint machine
  learning analysis of muon spectroscopy data from different materials},
  Journal of Physics: Conference Series 2164~(1) (2022) 012018.
\newblock \href {https://doi.org/10.1088/1742-6596/2164/1/012018}
  {\path{doi:10.1088/1742-6596/2164/1/012018}}.
\newline\urlprefix\url{https://dx.doi.org/10.1088/1742-6596/2164/1/012018}

\end{thebibliography}





\end{document}